\documentclass{aastex}
\usepackage{emulateapj5}
\usepackage{apjfonts}
\usepackage{epsf}
\usepackage{times}

\usepackage{graphics}

\newcommand{\be}{\begin{equation}}
\newcommand{\ee}{\end{equation}}
\newcommand{\bd}{\begin{displaymath}}
\newcommand{\ed}{\end{displaymath}}

\slugcomment{Received 2004 January 14; accepted 2004 March 8}

\shorttitle{Evidence for anisotropic motion of the clouds in BLRs}
\shortauthors{Cao X}

\begin{document}

\footnote{Email: cxw@shao.ac.cn}

\title{Evidence for anisotropic motion of the clouds in broad-line regions of BL Lacertae objects}

\author{Xinwu Cao}
\affil{Shanghai Astronomical Observatory, Chinese Academy of
Sciences, 80 Nandan Road, Shanghai, 200030, China}

\begin{abstract}

The masses of central massive black holes in BL Lac objects are
estimated from their host galaxy absolute magnitude $M_{\rm R}$ at
$R$-band by using the empirical relation between absolute
magnitude of host galaxy $M_{\rm R}$ and black hole mass $M_{\rm
bh}$. Only a small fraction of BL Lac objects exhibit weak
broad-line emission, and we derive the sizes of the broad-line
regions (BLRs) in these BL Lac objects from the widths of their
broad emission lines on the assumption of the clouds being
virilized in BLRs. It is found that the sizes of the BLRs in these
sources are usually 2-3 orders of magnitude larger than that
expected by the empirical correlation $R_{\rm BLR}-\lambda
L_{\lambda}$(3000\AA) defined by a sample of Seyfert galaxies and
quasars \citep{mj02}. We discuss a variety of possibilities and
suggest it may probably be attributed to anisotropic motion of the
BLR clouds in these BL Lac objects. If the BLR geometry of these
sources is disk-like, the viewing angles between the axis and the
line of sight are in the range of $\sim 2^\circ-12^\circ$, which
is consistent with the unification schemes.

\end{abstract}


\section{Introduction}

There is a tight correlation between the black hole mass $M_{\rm
bh}$ and the stellar dispersion velocity $\sigma$
\citep{fm00,g00}.  This tight correlation $M_{\rm bh}-\sigma$ is
widely used to estimate the central black hole masses of active
galactic nuclei (AGNs). Unfortunately, the stellar dispersion
velocity $\sigma$ is available only for a small fraction of AGNs.
\citet{md02} derived a very tight correlation between host galaxy
absolute magnitude $M_R$ at $R$-band and black hole mass $M_{\rm
bh}$.
\citet{cao03} used this relation $M_{\rm bh}-M_R$ to estimate the
central black hole masses of 29 BL Lac objects. The central black
hole masses of three sources in their sample have also been
measured from the stellar dispersion velocity \citep{f02,b03},
which agree well with the black hole masses estimated from the
host galaxy absolute magnitude $M_R$. \citet{o02} also found that
the black hole masses estimated from the host galaxy luminosity
are quite reliable for radio galaxies.

The BLR sizes $R_{\rm BLR}$ of broad-line ${\rm H}\beta$ are
measured by \citet{k00} for a sample of quasars and Seyfert
galaxies using reverberation mapping method. They found a tight
correlation between the BLR size $R_{\rm BLR}$ and optical
continuum luminosity $\lambda L_{\lambda}$. Using the width of the
broad emission line and measured  BLR size $R_{\rm BLR}$, they
estimated the central black hole masses of the sources in their
sample assuming the clouds in BLRs to be virilized. For sources at
high redshifts, the emission of line ${\rm H}\beta$ is usually
unavailable. Instead, the width of the line Mg\,{\sc ii} can be
used to estimate the central black hole masses \citep{mj02}.
Mg\,{\sc ii} is a low-ionization line as ${\rm H}\beta$, so that
Mg\,{\sc ii} is expected to be produced in the same region as
${\rm H}\beta$, which is supported by the tight correlation
between the FWHM of Mg\,{\sc ii} and ${\rm H}\beta$ ($V_{\rm
FWHM}$({\rm Mg\,{\rm\sc ii}})$\sim V_{\rm FWHM}({\rm H}\beta$))
found by \citet{mj02}. It is therefore reasonable to expect that
Mg\,{\sc ii} and ${\rm H}\beta$ are produced in the same region.
Using the same sample and the BLR sizes measured by \citet{k00},
\citet{mj02} obtained a correlation between $R_{\rm BLR}$ and the
monochromatic continuum luminosity at 3000 {\AA}, which is useful
for estimate of black hole masses of the sources at high redshifts
with only Mg\,{\sc ii} emission line profiles.

For most AGNs, their BLR sizes have not been measured directly by
reverberation mapping method and the empirical relation $R_{\rm
BLR}-\lambda L_{\lambda}$ is used to derive $R_{\rm BLR}$.
Combining the line width, the central black hole masses can be
estimated by assuming the motion of clouds in BLRs to be
virilized. The estimated black hole masses depend sensitively on
the velocity of the clouds in BLRs ($\propto V_{\rm BLR}^2$). The
broad-line width is mainly governed by the component of the cloud
velocity $V_{\rm BLR}$ projected to the line of sight. If the
motion of BLR clouds is anisotropic (e.g., disk-like BLR
geometry), the estimate of black hole mass becomes complicated
(e.g., Jarvis \& McLure, 2002). \citet{md01} argued that BLRs in
some AGNs have disk-like geometry. If the disk-like BLR geometry
is indeed present, the observed broad-line width depends
sensitively on the orientation of disk axis and the orientation o
is therefore crucial in the estimate of black hole mass from its
broad-line width. In the unification schemes, the jets of BL Lac
objects are supposed to be inclined at small angles with respect
to the line of sight (see Urry \& Padovani, 1995 for a review), so
the broad-line profiles will be significantly narrowed if
disk-like BLRs are present perpendicular to the jets, which can be
used to test the geometry of BLRs in BL Lac objects. In this
paper, we use the observed broad emission line widths and black
hole masses derived from host galaxy luminosity to test the
geometry of BLRs in BL Lac objects if the black hole masses can be
estimated by independent method. The cosmological parameters
$\Omega_{\rm M}=0.3$, $\Omega_{\Lambda}=0.7$, and $H_0=70~ {\rm
km~s^{-1}~Mpc^{-1}}$ have been adopted in this paper.

\section{Black hole mass}

In order to estimate the central black hole masses of BL Lac
objects, we use the relation between host galaxy absolute
magnitude $M_R$ at $R$-band and black hole mass $M_{\rm bh}$
proposed by \citet{md04}
 \be
 \log_{10}(M_{\rm bh}/M_{\odot})=-0.50(\pm0.02)M_R -2.75(\pm0.53).
 \label{mrmbh}\ee
There are different surveys on host galaxies of BL Lac objects
(e.g., Urry et al. (2000); Pursimo et al. (2002); Nilsson et al.
(2003)).
In this paper, we search the literature for all BL Lac objects
with both measured host galaxies and broad emission line profiles.
As only a small fraction of BL Lac objects exhibit broad-line
emission, we finally obtain a sample of 16 BL Lac objects. All
data collected are list in Table 1, and the derived parameters of
these sources are listed in Table 2.  The apparent magnitudes of
the host galaxies at $R$-band listed in Table 1 are galactic
extinction and $K$-corrected. Only upper limits on host galaxy
luminosity are available for 9 sources in this sample. We estimate
the central black hole masses of these BL Lac objects from their
host galaxy luminosity, and their broad-line profiles are used to
explore their BLR geometry.  The black hole masses of two sources
in  our sample have been measured from the stellar dispersion
velocity $\sigma$, which gives $\log(M_{\rm
bh}/M_{\odot})$=8.65(0521$-$365), 8.51(1807$+$698) \citep{b03},
and 8.90(1807$+$698) \citep{f02}, respectively. The black hole
masses of these two sources derived from its host galaxy
luminosity are $\log(M_{\rm bh}/M_{\odot})$=8.56(0521$-$365),
8.88(1807$+$698) (see Table 2). We can conclude that the black
hole masses derived from host galaxy luminosity are quite
reliable.

\section{BLR sizes of BL Lac objects}

The BLR size can be derived from the full-width half-maximum
(FWHM) of the broad line \be R_{\rm BLR}={\frac {GM_{\rm bh}}{f^2
V_{\rm FWHM}^2}}, \label{rblrmbh} \ee if the central black hole
mass $M_{\rm bh}$ is available. The correction factor $f=1/(2\sin
i)$ for a pure disk-like BLR of which the axis is inclined to the
line of sight at an angle $i$ \citep{md01}, while $f=\sqrt{3}/2$
for the clouds moving at random inclinations \citep{wpm99,k00}.
Eleven sources in our sample have measured FWHM for broad-line
Mg\,{\sc ii}, and four for H$\alpha$ and another for H$\beta$,
respectively. The broad-line Mg\,{\sc ii} is expected to be
produced in the same region of the BLR, so we take $V_{\rm
FWHM}$(Mg\,{\rm\sc ii})=$V_{\rm FWHM}$({\rm H}$\beta$) for the
source of which only the line width of H$\beta$ is available.
Using the black hole mass $M_{\rm bh}$ derived from the host
galaxy luminosity and velocity $V_{\rm FWHM}$, we can calculate
the BLR size by using Eq. (\ref{rblrmbh}) on the assumption of
isotropic motion of clouds in the BLR, i.e., $f=\sqrt{3}/2$ is
adopted. We convert the size $R_{\rm BLR}(\rm H\alpha)$ to $R_{\rm
BLR}(\rm H\beta)$ using relation (2) \be R_{\rm BLR}({\rm
H}\alpha)=1.19(\pm0.23)R_{\rm BLR}({\rm H}\beta)+13(\pm19)~~{\rm
lt-day} \ee in \citet{k00} for the sources of which only the line
widths of H$\alpha$ are available.

\section{The ionizing luminosity}

BL Lac objects exhibit non-thermal continuum emission, which is
believed to be dominated by the emission from the jets moving at
relativistic speed at small angles with respect to the  line of
sight \citep{br78}. The observed optical continuum emission is a
mixture of the emission from the disk and beamed emission from the
jet, which prevents us from measuring ionizing luminosity directly
from observed optical continuum emission for BL Lac objects (see
Urry \& Padovani, 1995 for a review and references therein).
Narrow-line emission is suggested  to be a good tracer for
ionizing luminosity of radio-loud AGNs \citep{rs91}. We use the
luminosity of the narrow-line {[O\,{\sc ii}]} at 3727~{\AA} to
estimate the optical ionizing luminosity. For those the emission
data of {[O\,{\sc ii}]} being unavailable, we estimate the
{[O\,{\sc ii}]} emission from other emission lines adopting the
line ratio proposed by \citet{f91}. The measured equivalent widths
of narrow-line {[O\,{\sc ii}]} corrected to the source frame
corresponding to the measured continuum emission are usually
around a few {\AA} or less than 1 {\AA} for the BL Lac objects of
which the narrow-line emission is detected (e.g., Stickel, Fried,
\& K\"uhr (1989; 1993); Lawrence et al. (1996); also see Table 1
for the sources in our sample). As the observed optical continuum
emission may be dominated by the beamed jet emission, the
equivalent width EW$_{\rm ion}$ of narrow-line {[O\,{\sc ii}]}
corresponding to the ionizing continuum emission from the
accretion disk should be larger than EW$_{\rm obs}$ measured
directly from the observed spectrum. The narrow-line emission is
isotropic and independent of the jet orientation. and we can
therefore use the narrow-line luminosity $L_{[\rm O\,{\rm\sc
II}]}$ to estimate the ionizing luminosity at 3000 {\AA} for these
BL Lac objects. Here we use the equivalent width of the line
{[O\,{\sc ii}]}: ${\rm EW}_{\rm ion}=10$ {\AA}, corresponding to
the ionizing continuum emission, which is suggest by  \citet{w99}
to estimate ionizing luminosity of radio-loud AGNs.

Using the same sample and the  BLR sizes derived by \citet{k00},
\citet{mj02} obtained a correlation between $R_{\rm BLR}^{\rm
emp}$ and the monochromatic continuum luminosity at 3000 {\AA} \be
R_{\rm BLR}^{\rm emp}=(25.2\pm3.0)\left({\frac {\lambda
L_{\lambda,3000}}{10^{37}{\rm W}}}\right)^{0.47\pm0.05},
\label{rblr3000}\ee which is useful especially for the sources at
high redshifts. In order to avoid confusion with the BLR sizes
derived from Eq. (\ref{rblrmbh}), we use $R_{\rm BLR}^{\rm emp}$
to represent the BLR sizes derived from the empirical relation
(\ref{rblr3000}).

\section{Results and discussion}

We plot the relation between the ionizing luminosity $\lambda
L_{\lambda}(3000)$ at 3000 {\AA} and  BLR size $R_{\rm BLR}$ in
Fig. \ref{fig1}. The BLR size $R_{\rm BLR}$ is derived from the
width of broad line Mg\,{\rm\sc ii} and black hole mass $M_{\rm
bh}$ on the assumption of isotropic motion of clouds in the BLR,
i.e., $f=\sqrt{3}/2$ is adopted in Eq. (\ref{rblrmbh}). We can
also derive the BLR size $R_{\rm BLR}^{\rm emp}$ using the
empirical relation (\ref{rblr3000}). If the motion of BLR clouds
is indeed isotropic, one may expect similar BLR sizes derived by
these two different methods. However, it is found that the sizes
of BLRs $R_{\rm BLR}$ in all these sources are $\sim$2-3 orders of
magnitude larger than $R_{\rm BLR}^{\rm emp}$ expected by relation
(\ref{rblr3000}).

We note that only upper limits on the masses of the black holes in
9 of all 16 sources. The BLR sizes may be over-estimated for these
9 sources, because of the BLR sizes $R_{\rm BLR}$  being derived
from the line widths and black hole masses (see Eq.
\ref{rblrmbh}). The black hole masses estimated from the host
galaxy luminosity for these 16 sources are in the range of $\sim
10^{8.6-10.4} M_{\odot}$ (see Table 1).  The deviations of the BLR
sizes $R_{\rm BLR}$ from $R_{\rm BLR}^{\rm emp}$ expected by
relation (\ref{rblr3000}) cannot be solely attributed to the
overestimate of black hole masses for those nine sources with
upper limits on galaxy luminosity, unless the black hole masses
have been overestimated by 2$-$3 orders of magnitude, i.e., the
realistic black hole masses should be in the range of $\sim
10^{6-8}$ $M_\odot$ for these sources, which seems impossible. It
will be more difficult to attribute such deviations to the
overestimate of black hole masses for those 7 sources with well
measured host galaxy luminosity.  The black hole mass of the
source $1807+698$ has been measured from its stellar dispersion
velocity $\sigma$ \citep{f02,b03}, which is consistent with our
estimate of the black hole mass $10^{8.88} M_{\odot}$. For this
source, its BLR size $R_{\rm BLR}$ derived from relation
(\ref{rblrmbh}) is about three orders of magnitude higher than
$R_{\rm BLR}^{\rm emp}$ predicted by relation (\ref{rblr3000})
between $R_{\rm BLR}^{\rm emp}$ and $\lambda L_{\lambda}(3000)$.

\figurenum{1}
\centerline{\includegraphics[angle=0,width=10.0cm]{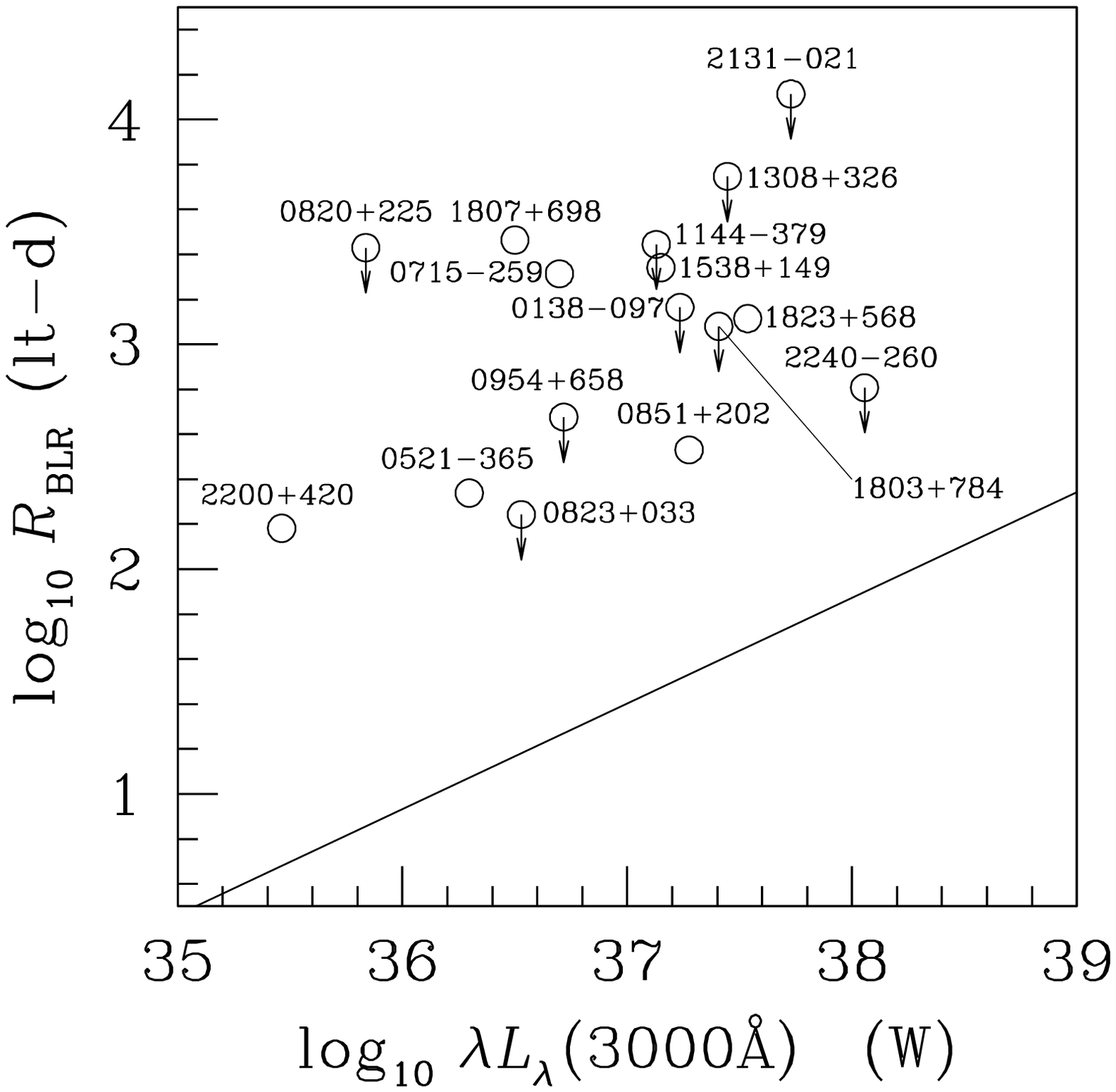}}
\figcaption{\footnotesize  The relation between the BLR size
$R_{\rm BLR}$ and the ionizing luminosity $\lambda
L_{\lambda}(3000)$ at 3000 {\AA}  derived from narrow-line
luminosity $L_{[{\rm O}_{\rm II}]}$. The line represents the
correlation $R_{\rm BLR}^{\rm emp}-\lambda L_{\lambda}(3000)$
defined by Seyfert 1 galaxies and quasars  \citep{mj02}. The
source names are labelled in the plot.\label{fig1}} \centerline{}


The ionizing luminosity at 3000 {\AA} is derived from the
luminosity of narrow-line [O\,{\sc ii}] adopting EW$_{\rm ion}$
=10 {\AA} \citep{w99}, which is in general consistent with the
optical spectroscopic observations on the radio quasars selected
from the Molonglo Quasar Sample (MQS), of which the equivalent
widths of [O\,{\sc ii}] in the source frame are in the range from
less than 1 {\AA} to more than 100 {\AA} with an average of 14.7
{\AA} \citep{b99}.  However, using a single value EW$_{\rm ion}$
=10 {\AA} is still a rough estimate, and may induce uncertainties
on the estimates of ionizing luminosity. In Fig. \ref{fig1}, we
find that the deviations of $R_{\rm BLR}$ from $R_{\rm BLR}^{\rm
emp}$ will not be solved on the isotropic BLR geometry assumption
even if a rather small  EW$_{\rm ion}$=1 {\AA} is adopted, i.e.,
the ionizing luminosity is an order of magnitude higher than the
present values. In Table 1, we have listed the measured equivalent
widths of [O\,{\sc ii}] in the source frame for the sources in our
present sample, which are in 0.2$-$9.2 {\AA}. Considering the
optical continuum emission contributed by the beamed emission from
the jets in these BL Lac objects, their equivalent widths
(EW$_{\rm ion}$) corresponding to the ionizing optical continuum
emission should be larger than the measured values listed in Table
1. This implies that the uncertainties on the estimates of the
ionizing optical continuum luminosity would not change the
conclusion on this point.  If photo-ionization of the gases in
narrow-line region by radiative shocks driven by the radio source
is important (e.g., Inskip et al. 2002), the central ionizing
luminosity should be lower than the present values, which would
lead to even larger deviations.

\citet{wz03} found that the BLR sizes of dwarf
AGNs are systematically larger than the prediction of $R_{\rm
BLR}-\lambda L_{\lambda}$ correlation defined by Seyfert galaxies
and quasars \citep{k00}. They suggested that the flat ionizing
spectra are in these dwarf AGNs as predicted by advection
dominated accretion flow (ADAF) models, and the BLRs in those
dwarf AGNs have lower ionization or/and lower density than those
in Seyfert 1 galaxies and quasars of \citet{k00}'s sample. The
sources in our present sample have brighter ionizing luminosity
than those dwarf AGNs. As most BL Lac objects do not exhibit any
line emission, all these 16 BL Lac objects have measured
broad-line profiles and narrow-line emission have relatively high
ionizing luminosity amongst all BL Lac objects \citep{cao02a}.
There is a critical accretion rate $\dot{m}_{\rm crit}$, and an
ADAF can exist only if accreting at a rate $\dot{m}<\dot{m}_{\rm
crit}$(e.g., Mahadevan, 1997), which leads to  an upper limit on
optical continuum emission from an ADAF for a given black hole
mass \citep{cao02a}. \citet{cao03} calculated optical spectra of
ADAF$+$SD(standard disk) systems and compared them with the
observed spectra, which suggests that only ADAFs themselves are
unable to produce such bright ionizing optical continuum emission
and standard thin disks should be present at least in the outer
regions of the disks for these BL Lac objects. The ionizing photos
are therefore mainly from the standard thin disk regions  in these
BL Lac objects, unlike the dwarf AGNs considered by \citet{wz03}.
The luminosity $\lambda L_{\lambda}(3000)$ of the sources in the
sample used to derive the correlation $R_{\rm BLR}-\lambda
L_{\lambda}(3000)$ is in the range of $\sim 10^{34-39}$ W
\citep{mj02}.  The ionizing luminosity of the BL Lac objects in
present sample at 3000 {\AA} are in the range of $\sim
10^{35.8-38}$ W, which is in the similar range as their sample.
The $R_{\rm BLR}-\lambda L_{\lambda}(3000)$ relation derived from
a sample of quasars and Seyfert galaxies should be valid for these
BL Lac objects, unless the physics of BLRs in these BL Lac objects
is significantly different from that in the sources of the sample
considered by \citet{mj02}.

We derive the BLR sizes of these BL Lac objects assuming the
motion of clouds to be isotropic, which may not be the case in
these sources. Our estimate of  BLR sizes may strongly be
overestimated due to anisotropic cloud motion, if the velocity
component projected to the line of sight is only a small fraction
of its real velocity. A most likely candidate for such anisotropic
motion of clouds is the clouds orbiting in a disk-like BLR (e.g.,
McLure \& Dunlop, 2001). For the clouds orbiting in a disk-like
BLR, the correction factor $f$ in Eq. (\ref{rblrmbh}) is $1/(2\sin
i)$, where  $i$ is the angle of the axis inclined to the line of
sight \citep{md01}. If this is the case, we can estimate the
inclination angle $i$ of these BL Lac objects assuming they indeed
to obey the correlation $R_{\rm BLR}-\lambda L_{\lambda}(3000)$
suggested by \citet{mj02}, i.e., the value of $f$ is estimated by
letting $R_{\rm BLR}=R_{\rm BLR}^{\rm emp}$. The derived results
are listed in Table 1. We find that the inclination angles are
around $\sim 2^\circ-12^\circ$ for these BL Lac objects.  There is
evidence that the velocity field of BLR is better described by a
combination of a random isotropic component, with  characteristic
velocity $V_{\rm r}$, and a component only in the plane of the
disk, with characteristic velocity $V_{\rm p}$ \citep{wb86}. In
this case, the observed FWHM will be given by \be V_{\rm
FWHM}=2(V_{\rm r}^2+V_{\rm p}^2\sin^2 i)^{1/2} \ee\citep{md01}, so
$f=0.5[(V_{\rm r}/V_{\rm p})^2+\sin^2 i]^{-1/2}$. If the random
isotropic component is important, i.e., $V_{\rm r}$ is comparable
with $V_{\rm p}$, then the term $(V_{\rm r}/V_{\rm p})^2$ cannot
be neglected and the derived inclined angle of the disk axis will
be less than that listed in Table 2. This is in general consistent
with the unification schemes that the jets of BL Lac objects are
inclined at small angles to the line of sight.

\acknowledgments

This work is supported by NSFC(No. 10173016; 10325314, 10333020)
and the NKBRSF (No. G1999075403). This research has made use of
the NASA/IPAC Extragalactic Database (NED), which is operated by
the Jet Propulsion Laboratory, California Institute of Technology,
under contract with the National Aeronautic and Space
Administration.

\begin{deluxetable}{ccccccccc}
\tabletypesize{\scriptsize} \tablecaption{Data of BL Lac objects}
\tablewidth{0pt} \tablehead{ \colhead{Source} & \colhead{Redshift}
& \colhead{$m_R$(host)} & \colhead{References} &
\colhead{FWHM(Mg\,{\sc ii})} & \colhead{References} &
\colhead{$\log~\lambda L_{\lambda}$} & EW of [O\,{\sc ii}] &
\colhead{References} } \startdata
 0138$-$097 &  0.733 & $>$18.38 & 1& 4842   &  4  & 37.23  & 0.52 & 4\\
 0521$-$365 &  0.055 & ~~14.35  & 1& ~~~3000$^{\rm a,b}$   & 5  & ~36.30$^{\rm d}$ &  & 11\\
 0715$-$259 &  0.465 & ~~16.76 & 1& 5200 & 6 & 36.70 & 1.3 & 6\\
 0820$+$225 &  0.951 & $>$19.38  &1  & 2995   &  7   & ~35.84$^{\rm e}$ &  & 7\\
 0823$+$033 &  0.506 & $>$19.04 & 1 & 5455 &  7  & 36.53 & 0.2 & 7\\
 0851$+$202 &  0.306 & ~~18.44  & 2& ~~2635$^{\rm c}$ & 8   & ~37.28$^{\rm f}$ & & 8\\
 0954$+$658 &  0.367 & $>$18.85 & 1 & ~~2079$^{\rm a}$ &  9  & 36.72 & 0.22 & 4\\
 1144$-$379 &  1.048 & $>$19.93 & 1 & 2492 &  8  & ~37.13$^{\rm e}$ & & 8\\
 1308$+$326 &  0.996 & $>$18.36  & 2,3& 4016 & 7 & ~37.45$^{\rm e}$ & & 7\\
 1538$+$149 &  0.605 & ~~18.73  & 1& 2411  &  7 & 37.15 & 0.44 & 7\\
 1803$+$784 &  0.684 & $>$19.15 & 1 & 3082 &  4  & ~37.41$^{\rm f}$ & & 9\\
 1807$+$698 &  0.051 & ~~13.54 &  1& ~~1326$^{\rm a}$ &  3   & 36.50 & 2.5 & 7\\
 1823$+$568 &  0.664 & ~~18.57  & 1 & 3952 &  4  & 37.54 & 1.9 & 7\\
 2131$-$021 &  1.285 & $>$18.50 & 1 & 3602 &  4   & 37.73 & 9.2 & 4\\
 2200$+$420 &  0.069 & ~~14.55 & 1& ~~4260$^{\rm a}$ & 10 & ~35.46$^{\rm f}$ &  & 7\\
 2240$-$260 &  0.774 & $>$20.22  &1 & 2753 &  7  & 38.06 & 1.4 & 7\\
 \enddata
 \tablenotetext{a}{FWHM of H$\alpha$.}
\tablenotetext{b}{The profile of broad-line H$\alpha$ in this
source is asymmetric with a red wing of FWHM $\simeq$
3000~km~s$^{-1}$ and a blue wing of FWHM $\simeq$ 1500~km~s$^{-1}$
\citep{s95}. Here, we conservatively take
FWHM$=$3000~km~s$^{-1}$.}
 \tablenotetext{c}{FWHM of H$\beta$,}
 \tablenotetext{d}{the equivalent width of [O\,{\sc ii}] is not given,}
 \tablenotetext{e}{the flux of [O\,{\sc ii}] is converted from Mg\,{\sc ii} by using the ratio given in \citet{f91},}
  \tablenotetext{f}{the flux of [O\,{\sc ii}] is converted from [O\,{\sc iii}] by using the ratio given in \citet{f91},}
 \tablecomments{Column (1):
source name; Column (2): redshift; Column (3): galactic extinction
and $K$-corrected $R$-band magnitude of the host galaxy; Column
(4): references for $R$-band magnitude of the host galaxy; Column
(5): FWHM of broad-line Mg\,{\sc ii} (km~s$^{-1}$); Column (6):
references for FWHM(Mg\,{\sc ii}); Column (7): ionizing luminosity
at 3000~$\rm\AA$ (W) estimated from $L_{[\rm O{\rm II}]}$; Column
(8): measured equivalent width of  [O\,{\sc ii}] corrected to the
source frame ($\rm\AA$); Column (9): references for $L_{[\rm O{\rm
II}]}$.} \tablerefs{(1) \citet{u00}; (2) \citet{p02}; (3)
\citet{n03}; (4) \citet{rs01}; (5) \citet{s95}; (6) \citet{c03};
(7) \citet{s93}; (8) \citet{s89}; (9) \citet{l96}; (10)
\citet{c96}; (11) \citet{t93}. }
\end{deluxetable}

\begin{deluxetable}{ccccc}
\tabletypesize{\scriptsize} \tablecaption{Derived parameters of BL
Lac objects} \tablewidth{0pt} \tablehead{ \colhead{Source}
&\colhead{$\log~M_{\rm bh}/M_\odot$ }  & \colhead{$\log~R_{\rm
BLR}$ } & \colhead{$f$}  & \colhead{$i$ } } \startdata
 0138$-$097 & ~$<$9.70  & $<$3.16 & $<$5.8 & $>$4.9$^\circ$ \\
 0521$-$365 & ~~~8.56 & ~~2.34 & ~~4.2 & ~~6.9$^\circ$ \\
 0715$-$259 &  ~~~9.91 & ~~3.31 & ~~9.2 & ~~3.1$^\circ$\\
 0820$+$225 &   ~$<$9.55 & $<$3.43 & $<$16.8 & $>$1.7$^\circ$\\
 0823$+$033 &   ~$<$8.88 &  $<$2.24 & $<$2.9 & $>$9.8$^\circ$\\
 0851$+$202 &  ~$<$8.78 & $<$2.78 & $<$3.6 & $>$7.9$^\circ$\\
 0954$+$658 &  ~$<$8.56 & $<$2.68 & $<$4.8 & $>$5.9$^\circ$\\
 1144$-$379 &  ~$<$9.40 & $<$3.44 & $<$8.5 & $>$3.4$^\circ$\\
 1308$+$326 &   $<$10.12 & $<$3.75 & $<$10.1 & $>$2.8$^\circ$\\
 1538$+$149 & ~~~9.27 & ~~3.34  & ~~7.4 & ~~3.8$^\circ$\\
 1803$+$784 &  ~$<$9.22 &  $<$3.08 & $<$4.8 & $>$6.0$^\circ$\\
 1807$+$698 &  ~~~8.88 & ~~3.46  & ~12.2 & ~~2.3$^\circ$\\
 1823$+$568 &   ~~~9.47 & ~~3.11  & ~~4.7 & ~~6.2$^\circ$\\
 2131$-$021 &  $<$10.39 &  $<$4.11 & $<$13.3 & $>$2.2$^\circ$\\
 2200$+$420 &  ~~~8.71 & ~~2.18 & ~~5.5 & ~~5.2$^\circ$\\
 2240$-$260 &  ~$<$8.85 &  $<$2.81 &  $<$2.5 & $>$11.7$^\circ$\\
 \enddata
 \tablecomments{Column (1):
source name;  Column (2): black hole mass; Column (3): BLR size
for Mg\,{\sc ii} (light-day); Column (4): derived correction
factor $f$ for BLR geometry; Column (5): inclination angle of the
jet with respect to the line of sight.}

\end{deluxetable}


\end{document}